\documentclass[12pt]{article}
\usepackage{amsmath,amsfonts,latexsym}
\usepackage{amssymb}
\usepackage{graphics}

\newcommand{\nc}{\newcommand}
\nc{\tr}{{\vartriangle}} \nc{\vth}{{\vartheta}}
 \nc{\bt}{{\beta}}
\nc{\dl}{{\delta}} \nc{\Dl}{{\tr}}
 \nc{\p}{{\psi}}
\nc{\gm}{{\gamma}} \nc{\Gm}{{\Gamma}} \nc{\sg}{{\sigma}}
\nc{\ve}{{\varepsilon}} \nc{\ch}{{\cal H}} 
\nc{\cf}{{\cal F}}
\nc{\cp}{{\cal P}}
 \nc{\td}{\tilde}

\newtheorem{lemma}{Lemma}[section]

\newtheorem{corollary}[lemma]{Corollary}

\newtheorem{remark}[lemma]{Remark}
\numberwithin{equation}{section}

\begin{document}

\title
{Truncated Stochastic Approximation  with Moving Bounds: Convergence}
   \author{Teo Sharia}
\date{}
\maketitle
\begin{center}
{\it
Department of Mathematics \\Royal Holloway,  University of London\\
Egham, Surrey TW20 0EX \\ e-mail: t.sharia@rhul.ac.uk }
\end{center}


\begin{abstract}
 In this paper we propose a  wide class of truncated stochastic approximation procedures.
These procedures have  three main characteristics:  truncations  with random moving bounds, 
 a matrix valued random step-size sequence,  and  a dynamically changing random regression function.
 We establish convergence and consider several examples to illustrate  the results.
 
 \end{abstract}


\begin{center}
Keywords: {\small 
Stochastic approximation,   Recursive estimation,  Parameter estimation}
\end{center}

\section{Introduction}

Stochastic approximation (SA) introduced by  Robbins and Monro  in 1951 (\cite{RM}) 
was created to locate a root of an unknown function when  only noisy measurements  of  
the function  can  be observed.  SA  quickly became  very popular, 
resulting in interesting new developments and numerous applications across a  wide range of disciplines. 
Comprehensive surveys of the SA technique  including  some  recent developments can be found 
in \cite{Ben}, \cite{Bor}, \cite{Kush1}, \cite{Kush2}, \cite{Lai}.

 In this paper we propose a  wide class of truncated SA procedures with moving random bounds.
 While we believe that  the proposed class of procedures  will find its way to  a wider range
 of applications,  the main motivation is to accommodate  applications to  parametric statistical estimation theory.
 Our  class of SA procedures  has three main characteristics:  truncations  with random moving bounds, 
 a matrix-valued random step-size sequence,  and  a dynamically changing random regression function.  
 
 To introduce the main idea, let us first consider  the  classical problem of 
 finding a unique zero, say $z^0$,  of  a real valued function $R(z):  \mathbb{R}  \to \mathbb{R}$ when only noisy 
 measurements  of $R$ are available. To estimate $z^0$, consider a sequence defined recursively   as  
\begin{equation}\label{S}
Z_t=
\big[~
Z_{t-1}+\gm_t \left(R(Z_{t-1})+\ve_t\right)\big]_{\alpha_t}^{\beta_t},   \qquad t=1,2,\dots
\end{equation}
where $\ve_t$ is a sequence of zero-mean random variables and  $\gamma_t$ is a 
deterministic sequence of positive numbers. Here
$\alpha_t$ and $\beta_t$ are random variables with $-\infty \le \alpha_t \le \beta_t\le \infty$
and $[v]_{a}^{b}$ is the truncation
operator, that is,
$$
[v]_{a}^{b} =
\begin{cases}
   a & \text{if} \;\; v<a,  \\
    v& \text{if} \;\;  a\le v\le  b,\\
    b & \text{if} \;\;v>b .
\end{cases}
$$

We assume that the  truncation sequence  $[\alpha_t,\beta_t]$  contains $z^0$ for large values of $t$. 
For example,  if  it is known that  $z^0$ belongs to  $(\alpha,\beta)$, with   $-\infty \le \alpha \le \beta\le \infty$, 
 one can  consider truncations with expanding bounds  to avoid possible singularities  at the endpoints of
 the interval. That is, we can take $[\alpha_t,\beta_t]$ with some  sequences  $\alpha_t \downarrow\alpha$ and
 $\beta_t\uparrow\beta$.  Truncations with expanding bounds  may also be useful to  overcome standard 
 restrictions on growth of the corresponding functions.

 The most interesting case arises when the truncation interval $[\alpha_t, \beta_t]$  represents our auxiliary
knowledge  about $z^0$   at step $t$,  which is incorporated into  the procedure through the truncation operator.
Consider for example a  parametric statistical model. Suppose that   
  $X_1, \dots, X_t$    are    independent and identically distributed random variables
    and $f(x, \theta)$ is    the  common probability density  function 
  (w.r.t. some $\sigma$-finite measure) 
   depending on an unknown parameter $\theta \in \mathbb{R}^m $.   Consider the recursive estimation procedure for $\theta$ defined by
\begin{equation} \label{reciid}
\hat\theta_t=\hat\theta_{t-1}+
\frac  {1}{t} i(\hat\theta_{t-1})^{-1}~\frac{{f'}^T(X_t, \hat\theta_{t-1})}{f(X_t, \hat\theta_{t-1})},
~~~~~~~~~t\ge 1.
\end{equation}
where  $f'$ is the row-vector of partial derivatives of $f$ w.r.t. the components of $\theta$,
 $i(\theta)$ is the one-step Fisher information matrix,
 and $\hat\theta_0\in {\mathbb{R}}^m$ is some initial value.  
This estimator was introduced   in \cite{Sakr} and studied in  \cite{Fab},  \cite{Khas} and \cite{Pol}.   In particular, it has been shown
that under certain conditions the recursive estimator $\hat\theta_t$ is asymptotically equivalent to the maximum likelihood estimator, i.e., it is  consistent and asymptotically efficient.  
The analysis  of   \eqref{reciid}
can be conducted  by rewriting it in the form of stochastic approximation.  Indeed, in the case of \eqref{reciid},  let us fix 
$\theta$ and  let $\gamma_t=1/t$,
$$
R(z)=i(z)^{-1} E^{\theta}\left\{ \frac{{f'}^T(X_t, z)}{f(X_t, z)}\right\}
~~~ \mbox{and} ~~~ 
\ve_t=i(\hat\theta_{t-1})^{-1} \left(\frac{{f'}^T(X_t, \hat\theta_{t-1})}{f(X_t, \hat\theta_{t-1})} - R(\hat\theta_{t-1})\right)
$$ 
($E^{\theta}$ is  expectation w.r.t. $f(x, \theta)$).
Then, under the usual regularity assumptions, $R(\theta)=0$ and $\ve_t$ is a martingale difference
(w.r.t. the filtration ${\cf}_{t}$ generated by the observations). So,  \eqref{reciid} is  a standard SA of  type \eqref{S} 
without truncations (i.e.,  in the one dimensional case,  $-\alpha_t =\beta_t = \infty$). 

However, the need of truncations may  naturally arise from various reasons. One obvious consideration 
is that the functions in the procedure  may only be defined for   certain values of the parameter.  In this case one would want 
the procedure to produce points only from this set. Truncations  may  also be useful when the  standard assumptions 
 such as restrictions on  the  growth rate of the relevant functions are not satisfied.   
  More importantly,  truncations may 
  provide a simple tool  to achieve an efficient use of   information available in the estimation process.
  This information can be auxiliary information about the parameters, e.g. a set, possibly time dependent, that is known to contain the value of the unknown parameter. Suppose for instance  that 
 a consistent (i.e., convergent), but not necessarily efficient auxiliary estimator $\tilde\theta_t$  is
available having a  rate $d_t$. Then one can consider a truncated procedure with  shrinking  bounds.
The idea is  to obtain asymptotically efficient estimator   by 
truncating  the recursive procedure in a neighbourhood of $\theta$
 with $ [\alpha_t, \beta_t]=[\tilde\theta_t-\delta_t, \tilde\theta_t+\delta_t],$
$\delta_t \to 0$. Such a procedure is obviously consistent
since $\hat\theta_t\in[\tilde\theta_t-\delta_t, \tilde\theta_t+\delta_t]$ and
$\tilde\theta_t\pm\delta_t \to \theta$.  However,   to construct an efficient estimator, care should be taken
to ensure that the truncation intervals  do not shrink to $\tilde\theta_t$ too rapidly,  for otherwise
 $\hat\theta_t$  will have the same asymptotic  properties as   $\tilde\theta_t$  (see \cite{Shar4} for
 details in the case of $AR$ processes).  Since this paper is concerned  with the convergence,
 details of this application is not discussed here. However,  since the procedures with shrinking bounds
 are particular cases of the general SA  procedure below (see \eqref {TSA}),   asymptotic distribution
 and efficiency  can be studied  in an unified manner using ideas  of  SA.  
 
 Note that the idea of truncations with moving bounds is  not new.  For example, an idea of truncations with shrinking 
 bounds goes back to \cite{Fab} and \cite{Khas}. Truncations with expanding bounds were considered in \cite{Andr}
 and also,  in the context of
 recursive parametric estimation, in \cite{Shar0} (see also \cite{Shar4}).  
 Truncations with adaptive truncation  sets of the Robbins-Monro SA   were 
 introduced   by Chen and Zhu in \cite{Chen1},  and  further explored  and extended  in    \cite{Chen2}, \cite{AMP},  \cite{Tadic1}, \cite{Tadic2}, \cite{Lel}.  
  The latter  algorithms are designed in such a way, that the procedure is pulled back to a certain pre-specified 
 point or a set, every time  the sequence leaves the truncation region.  As one can see  from \eqref{S}
 and \eqref {TSA}, truncation
 procedures considered in this paper  are quite different from the ones by  Chen and Zhu and 
  are similar to the  the ones introduced by Andrad\'ottir in \cite{Andr} (see Rematk \ref{Comp}).  A detailed comparison of these
  two different approaches is given in \cite{Andr}. 
 
 Let us now consider a discrete time stochastic processes   $X_1,  X_2, \dots $  with the
joint distribution depending on an unknown parameter $\theta \in \mathbb{R}^m $.
 Then one can consider the recursive estimator of   $\theta$ defined by
\begin{equation} \label{rec1}
\hat\theta_t=\hat\theta_{t-1}+
{\gamma_t(\hat\theta_{t-1})}\psi_t(\hat\theta_{t-1}),
~~~~~~~~~t\ge 1,
\end{equation}
where   $\psi_t(v)=\psi_t(X_1, \dots, X_t; v),$  $t=1,2,\dots, $
are suitably chosen functions which may, in general, depend on the  vector of all past and present random variables 
and have the  property that   the process $\psi_t(\theta)$ is  $P^\theta$- martingale difference, i.e.,
$E^\theta\left\{ \psi_t(\theta)\mid{\cal{F}}_{t-1}\right\}=0$ for each $t$. For example, 
  if $f_t(x,\theta)=f_t(x,\theta|X_1, \dots,X_{t-1})$ is the conditional probability density function 
of the observation $X_t$ given $X_1, \dots,X_{t-1},$   then one
can obtain a   likelihood type estimation procedure    by choosing
$\psi_t(v)=l_t(v)= f'_t(X_t,v)/f_t(X_t,v).$    Asymptotic behaviour of this type of procedures for non i.i.d. models
was studied by a number of authors, see e.g.,   \cite{Cam}, \cite{Eng}, \cite{Ljun2}, \cite{Shar} -- \cite{Shar3}.
Results in \cite{Shar3} show that   to obtain   an  estimator with asymptotically optimal properties, one has to 
consider a  state-dependent matrix-valued random step-size sequence. One possible choice is $\gamma_t(u)$
with the property
$$
\gamma_t^{-1}(v)-\gamma_{t-1}^{-1}(v)=E_{\theta}\{ \p_t(v)l^{T}_t(v)\mid
{\cf}_{t-1}\}
$$  
In particular, to obtain a  recursive procedure which is asymptotically equivalent to  
the maximum likelihood  estimator, one has to consider  $l_t(v)= f'_t(X_t,v)/f_t(X_t,v)$ and
$\gamma_t(v)=I_t^{-1}(v)$,   where   $I_t(v)$ is the conditional Fisher information  matrix (see  \cite{Shar3} for details).
 To rewrite \eqref{rec1}  in a SA form, let us assume that $\theta$ is an arbitrary but
fixed value of the parameter and  define
 $$
R_t(z)=E^{\theta}\left\{ \psi_t(X_t, z)\mid{\cf}_{t-1}\right\}
~~~ \mbox{and} ~~~ 
\ve_t(z)=\left(\psi_t(X_t, z) - R_t(z)\right).
$$ 
Obviously,  $R_t(\theta)=0$  for each $t$, and $\ve_t(z)$ is a martingale difference. 

Therefore, to  be able to study these procedures in an  unified  manner, one needs to consider a SA
of the following form 
$$                         
Z_t= \big[~
Z_{t-1}+\gm_t(Z_{t-1}) \big\{ R_t(Z_{t-1})+\ve_t(Z_{t-1}) \big\}\big]_{U_t},   \quad t=1,2,\dots
$$
where    $R_t(z)$ is  predictable   with the property 
that $R_t(z^0)=0$ for all $t$'s, $\gm_t(z)$ is a matrix-valued predictable step-size sequence,
 $U_t   \subset \mathbb{R}^m$  is  a random sequence of truncation sets, and  
$Z_0 \in \mathbb{R}^m$  is some starting value (see Section 2 for more details).

To summarise the above, the procedures introduced in this paper  have the  following features: (1) inhomogeneous  random functions $R_t$; (2) state dependent matrix valued random step sizes;  (3)  truncations with random and  moving 
(shrinking or expanding) bounds.
These are mainly motivated by parametric statistical 
applications. In particular,  (1) is required to include recursive parameter estimation procedures for  non i.i.d. models,
(2) is needed to guarantee asymptotic optimality and efficiency of  statistical estimation, (3) is required to accommodate
various different adaptive truncations, including the ones arising by auxiliary estimators. 
Also,  the convergence of  these procedures  is studied under very general conditions and the
 results might be of interest  even for the procedures without truncations (i.e., when
 $U_t=\mathbb{R}^m$) and with a  deterministic  and homogeneous regression function  
 $R_t(z)=R(z)$.

The paper is organised as follows.  In sections 2.2 we  prove two lemmas on the convergence. The analysis is based on
 the    method of   using  convergence sets 
 of nonnegative semimartingales.    The decomposition into negative and positive parts 
 in these lemmas turns out to be very useful in applications (see Example 3 in Section 2.4). 
 In section 2.3 we give several  corollaries  in the case  of state independent  scalar random step-size
 sequences. In section 2.4 we consider  examples.  Proofs of some technical parts are postponed to 
 Section 3.

\section{Convergence}



\subsection{Main objects and notation}\label{MON}

Let $(\Omega, ~ \cf,F=(\cf_t)_{t\geq 0}, ~P)$ be a stochastic basis satisfying the usual conditions. Suppose that 
for each $t=1,2, \dots$, we have 
$( {\cal{B}}  ( \mathbb{R}^m)  \times \cf )$-measurable functions
$$
\begin{array}{cl}
 R_t(z)= R_t(z,\omega) &:\mathbb{R}^m \times  \Omega   \to    \mathbb{R}^m  \\
 \ve_t(z)=\ve_t(z,\omega) &:\mathbb{R}^m \times  \Omega   \to    \mathbb{R}^m      \\
  \gamma_t(z)=\gamma_t(z,\omega)&:\mathbb{R}^m \times  \Omega   \to    \mathbb{R}^{m\times m}    
\end{array}
$$
such that 
 for each $z\in  \mathbb{R}^m$,  the   processes    $R_t(z) $  and  $\gamma_t(z)$ are  predictable, i.e.,
 $R_t(z) $  and  $\gamma_t(z)$ are $\cf_{t-1}$ measurable for each $t$.
 Suppose also that  
 for each $z\in  \mathbb{R}^m$,  the process $\ve_t(z) $   is a martingale difference, i.e., $\ve_t(z) $ 
 is  $\cf_{t}$ measurable and  $E\left\{\ve_t(z)\mid{\cal{F}}_{t-1}\right\}=0$.
 We also assume that 
$$
R_t(z^0)=0
$$  
for each $ t=1, \dots $, where  $z^o \in   \mathbb{R}^m$   is  a non-random vector. 
\medskip

Suppose that $h=h(z)$ is a real valued function  of
 $ z \in {{\mathbb{R}}}^m$.  We denote by $ h'(z)$  the row-vector
 of partial derivatives
of $h$ with respect to the components of $z$, that
is,
 $$
  h'(z)=\left(\frac{{\partial}}{{\partial} z_1} h(z), \dots,
 \frac{{\partial}}{{\partial} z_m} h(z)\right).
 $$
Also, we denote by  $h''(z)$ the  matrix of second partial derivatives.
 The $m\times m$ identity matrix is denoted by ${{\bf 1}}$.

Let    $U \subset    \mathbb{R}^m$ is a closed convex set  and define a truncation  operator    as a function
$\big[z\big]_{U} : \mathbb{R}^m  \longrightarrow \mathbb{R}^m$, such that
$$
\big[z\big]_{U}=\begin{cases}
    z & \text{if} \;\; z\in U \\
     z^* & \text{if} \;\; z\notin U,
\end{cases}
$$
where $z^*$ is a point in 
$U$,  that minimizes   the distance  to $z$.

Suppose that   $z^o \in   \mathbb{R}^m$.  We say that a random sequence   of sets $U_t =U_t(\omega)$ 
($ t=1,2, \dots $)  from $\mathbb{R}^m $  is {\em admissible}  for  $z^o$  if
\begin{description}
\item[$\bullet$ ] 
for each $t$ and  $\omega,$    $U_t(\omega)$   is a   closed convex  subset  of $ \mathbb{R}^m$;
 \item[$\bullet$ ] 
 for each   $t$ and $z \in   \mathbb{R}^m$, the truncation $\big[z\big]_{U_t}$  is  $ {\cal{F}}_{t}$ measurable; 
 
  \item[$\bullet$ ]  $z^o\in U_t$ eventually, i.e.,  for almost all  $\omega$ there exist  $t_0(\omega)<\infty$
 such that $z^o\in U_t(\omega)$ whenever $t >t_0(\omega)$.
 \end{description}

\medskip
Assume that $Z_0 \in \mathbb{R}^m$  is some starting value and  consider the procedure 
\begin{equation}\label{TSA}
Z_t=
\big[~
Z_{t-1}+\gm_t(Z_{t-1}) \Psi_t(Z_{t-1})\big]_{U_t},   \quad t=1,2,\dots
\end{equation}
were  $\Psi_t(z)=R_t(z)+\ve_t(z)$,  $U_t $   is {admissible}   for  $z^o$,   $R_t(z) $, $\ve_t(z)$, $\gm_t(z)$ are random fields  defined above,
\begin{equation}\label{GTSA1}
 E\left\{\Psi_t(Z_{t-1})\mid{\cal{F}}_{t-1}\right\}=R_t(Z_{t-1}),
  \end{equation}
\begin{equation}\label{GTSA2}
 E\left\{\ve_t^T(Z_{t-1})\ve_t(Z_{t-1})\mid{\cal{F}}_{t-1}\right\}= 
 \left[E\left\{\ve_t^T(z)\ve_t(z)\mid{\cal{F}}_{t-1}\right\} \right] _{z=Z_{t-1}},
 \end{equation}
 and the conditional expectations   \eqref{GTSA1} and \eqref{GTSA2}  are assumed to be finite.
 \medskip
%
\begin{remark}\label{disint}  Note that  \eqref{GTSA1}  in fact means that the sequence $\ve_t(Z_{t-1})$ is a martingale difference.
Conditions  \eqref{GTSA1} and \eqref{GTSA2} obviously hold if, e.g.,  the  measurement errors $\ve_t(u)$ are independent 
random variables, or if they  are state independent. In general,  
since we assume that all  conditional  expectations are calculated  as integrals w.r.t. corresponding regular conditional probability measures (see the convention below), these conditions can be checked using disintegration formula (see, e.g.,  Theorem 5.4 in \cite{Kall}).  
\end{remark}
\noindent
{\bf \em Convention.}

\noindent
$\bullet$
 {\em Everywhere in the present work
  convergence and all relations between random
variables are meant with probability one w.r.t. the measure
$P$ unless specified otherwise. \\
{$\bullet$} A sequence of random
variables $(\zeta_t)_{t\ge1}$ has some property  {\underline {{\bf \em eventually}}} if for
every $\omega$ in a set $\Omega_0$ of $P$ probability 1, the realisation 
 $\zeta_t(\omega)$  has this property for all $t$ greater than some
$t_0(\omega)<\infty$.}\\
{$\bullet$} We assume that all conditional expectations are calculated  as integrals w.r.t. corresponding regular conditional probability measures.\\
{$\bullet$}
{\em We will  also assume that    the $\inf_{z\in U} h(z)$ of a real valued function $h(z)$ is  $1$
whenever $U=\emptyset$.}

%


 \subsection{Convergence Lemmas}\label{CSL}

\begin{lemma}\label{SL} 
 Let  $Z_t$  be a  process  defined by \eqref{TSA}, \eqref{GTSA1}  and \eqref{GTSA2}, with an admissible for $z^0 \in  \mathbb{R}^m$  
 truncation sequence  $U_t$.  
 Let    $V(u)  :  \mathbb{R}^m\longrightarrow \mathbb{R} $
be  a real valued  nonnegative function  having  continuous and bounded partial second derivatives. Denote 
$$
\tr_t=Z_t-z^0
$$
and suppose that the following conditions are satisfied.
\begin{description}
 \item[(L)] 
$$
 V\Big(\tr_t\Big) \le  V\Big(\tr_{t-1}+\gm_t(Z_{t-1}) \Psi_t(Z_{t-1})\Big)
$$
eventually.
 \item[(S)]
\begin{equation}\label{N+conv}
\sum_{t=1}^\infty (1+V(\Dl_{t-1}))^{-1} \left[{\cal
N}_t(\Dl_{t-1})\right]^+ < \infty, \qquad
 P\mbox{-a.s.}
\end{equation}
where
\begin{eqnarray}
{\cal N}_t(u)&= & V'(u) \gm_t(z^o+u)R_t(z^o+u)\nonumber\\
&&
+\frac12\sup_{v} \|V''(v)\| E
\left\{\|\gm_t(z^o+u) \Psi_t(z^o+u)\|^2 \mid
{\cf}_{t-1}\right\}. \nonumber
\end{eqnarray}
\end{description}
Then $V(Z_t - z^o)$ converges  ($P$-a.s.) to a finite limit   for any  initial value $Z_0$.
Furthermore, 
\begin{equation}\label{N-conv}
\sum_{t=1}^\infty \left[{\cal
N}_t(\Dl_{t-1})\right]^- < \infty, \qquad
 P\mbox{-a.s.}
\end{equation}
\end{lemma}

 {\bf Proof.}  As always (see the convention in  \ref{MON}), convergence and all relations between random
variables are meant with probability one w.r.t. the measure
$P$ unless specified otherwise.

From condition (L), using  the Taylor expansion,
\begin{eqnarray}
V(\Dl_t) \le V(\Dl_{t-1})+ 
V'(\Dl_{t-1})\gm_t(z^o+\Dl_{t-1})
\Psi_t(z^o+\Dl_{t-1})  \nonumber\\
 +\frac 12 \left[\gm_t(z^o+\Dl_{t-1})
\Psi_t(z^o+\Dl_{t-1})\right]^T V''(\tilde
\Dl_{t-1}) \gm_t(z^o+\Dl_{t-1}) \Psi_t(z^o+\Dl_{t-1}),
\nonumber
\end{eqnarray}
where $\tilde \Dl_{t-1}\in \mathbb{R}^m$   is ${\cal{F}}_{t-1}$-measurable. 
Using \eqref{GTSA1}  and \eqref{GTSA2} and   taking the conditional
expectation w.r.t. $ {\cf}_{t-1}$ yields
\begin{equation}\label{previ} 
E\left\{V(\Dl_t)  \mid {\cal{F}}_{t-1}\right\} \le
V(\Dl_{t-1})+ {\cal N}_t(\Dl_{t-1}).
\end{equation}
Using the obvious decomposition $ {\cal N}_t(\Dl_{t-1})= {[{\cal
N}_t(\Dl_{t-1})]}^+ - {[{\cal N}_t(\Dl_{t-1})]}^-, $  we  can write
$$
 {\cal N}_t(\Dl_{t-1}) =
\left(1+V(\Dl_{t-1}) \right)^{-1}[{\cal
N}_t(\Dl_{t-1})]^+ \left(1+V(\Dl_{t-1}) \right) - [{\cal N}_t(\Dl_{t-1})]^-
$$
$$
=B_t\left(1+V(\Dl_{t-1}) \right)-  [{\cal N}_t(\Dl_{t-1})]^-.
$$
where
$$
B_t=\left(1+V(\Dl_{t-1}) \right)^{-1}[{\cal
N}_t(\Dl_{t-1})]^+.
$$ 
Hence  \eqref{previ}  implies that 
\begin{equation}\label{V}
E\left\{V(\Dl_t)  \mid {\cal{F}}_{t-1}\right\} \le
V(\Dl_{t-1})(1+B_t)+B_t- [{\cal N}_t(\Dl_{t-1})]^-,
\end{equation}
 eventually and,  by \eqref{N+conv},
\begin{equation}\label{B}
\sum_{t=1}^\infty B_t< \infty.
\end{equation}
According to the Robbins-Siegmund Lemma (see e.g.,  \cite{Rob2})
 inequalities \eqref{V} and \eqref{B}  imply that \eqref{N-conv} holds
and $V(\Dl_{t})$  converges to some finite limit.
$ \diamondsuit $

\vskip+0.2cm

Everywhere  below, we assume that    the $\inf_{u\in U} v(u)$ of a function $v(u)$ is  $1$
whenever $U=\emptyset$.

%

\begin{lemma} \label{CL} 
Suppose that   $V(Z_t - z^o)$ converges  ($P$-a.s.) to a finite limit   for any  initial value $Z_0$, where $Z_t$ and
$V$ are defined  in Lemma  \ref{SL},  and      \eqref {N-conv}    holds. Suppose also that    for each   $\ve\in (0, 1),$ 
\begin{equation}\label{VV}
\inf_{\stackrel{ \|u\| \geq \ve}{z^0+u \in U_t}} V(u)> \delta >0
\end{equation}
eventually,
for some $\delta$. Suppose also that 
\begin{description}
 \item[(C)]  For each $\ve\in (0, 1),$
$$
\sum_{t=1}^\infty \inf_u
\left[{\cal N}_t(u)\right]^-=\infty,    \qquad P\mbox{-a.s.}
$$
where the   infimum is taken over the set 
$\{u: {\ve \le V(u) \le {1/\ve}}; ~ {z^0+u \in U_{t-1}}\}$.
\end{description}
\end{lemma}
Then   $Z_t \to z^o \;\;
(P$-a.s.),  for any  initial value $Z_0$.

\medskip


 {\bf Proof.}  As always (see the convention in \ref{MON}), convergence and all relations between random
variables are meant with probability one w.r.t. the measure
$P$ unless specified otherwise. Suppose that 
   $V (\Dl_{t}) \to r \ge 0$  and  there exists a  set $A$ with $P (A)>0,$
such that
$r>0$ on $A$.  Then there
exists $\ve > 0$ and
(possibly random) $t_0,$ such that  if $t\ge t_0$, $\ve \le V(\Dl_{t-1}) \le {1/\ve}
$ on $A$. Note also that $z^o+\Dl_{t-1} =Z_{t-1}\in U_{t-1}$.  By
 {\bf {(C)}}, these would  imply that
$$
\sum_{s=t_0}^{\infty} [{\cal N}_s(\Dl_{s-1})]^- \ge
\sum_{s=t_0}^{\infty} \inf_u
\left[{\cal N}_s(u)\right]^-=\infty
$$
on the set $A$,
where the infimums  are taken  over the sets specified in  condition (C).  This contradicts  \eqref{N-conv}. Hence, $r=0$   and so,
 $ V(\Dl_{t}) \to  0$. Now, $\Dl_{t} \to  0$ follows from \eqref{VV} by contradiction.  Indeed, suppose that  $\Dl_{t} \not\to  0$ on a set, say $B$ of positive probability. Then, for any fixed $\omega$ from this set,  
  there would exist a sequence  $t_k \to \infty$  such that
 $\|\Dl_{t_k}\| \ge \ve $ for some $\ve>0,$ and \eqref{VV} would imply that
  $ V(\Dl_{t_k}) >  \delta>0$ for large $k$-s, which contradicts the  $P$-a.s.  convergence   $ V(\Dl_{t}) \to  0$.
  $ \diamondsuit $

\vskip+0.2cm


\subsection{Sufficient conditions}

Everywhere in this subsection  we   assume that $\gamma_t$ is state independent  
(i.e., constant w.r.t. $z$)  non-negative scalar predictable process. 
\begin{corollary}\label{SqS} 
 Let  $Z_t$  be a  process  defined by \eqref{TSA}, \eqref{GTSA1}  and \eqref{GTSA2}, with an admissible for $z^0 \in  \mathbb{R}^m$  
 truncation sequence  $U_t$.  
 Suppose also that $\gamma_t$ is  a  non-negative predictable scalar process and
\begin{description}
 \item[(C1)] \begin{equation}\label{N+}
 \sup_{z \in U_{t-1}} ~
\frac { \left[
2( z-z^o)^TR_t(z) +  \gm_t E \left\{ \|\Psi_t(z)\|^2 \mid{\cf}_{t-1}\right\} \right]^+}
  {1+\| z-z^o\|^2} \le q_t 
  \end{equation}
eventually, where
 $$
  \sum_{t=1}^{\infty}   {q_{t}}{\gamma_t} <\infty, \qquad P\mbox{-a.s.}
$$
\end{description}
 Then $\|Z_t-z^0\|$ converges ($P$-a.s.) to a finite limit.
\end{corollary}

 {\bf Proof.}   Let us show that the conditions of Lemma \ref{SL} are satisfied 
   with $V(u)=u^Tu=\|u\|^2$ and  the step-size sequence  $\gamma_t(z)=\gamma_t {\bf  I}$.
  Since    $z^0 \in U_t$ for large $t$-s,  the  definition of the truncation  (see \ref{MON})  implies that  
  $$
  \|Z_t - z^0 \| \le 
\left \| Z_{t-1}+\gm_t \Psi_t(Z_{t-1}) - z^0\right \|,   
$$
eventually. Therefore (L) holds.   Then, $V'(u)=2u^T$   and $V''(u)= 2 {\bf  I},$   and so, for 
the process ${\cal N}_t(u)$ in  \eqref{N+conv} we have 
\begin{equation}\label{NN}
{\cal N}_t(u)= 2u^T \gm_t R_t(z^o+u)+ \gm_t^2E\left\{ \| \Psi_t(z^o+u) \|^2 \mid {\cf}_{t-1} \right\} 
 \end{equation}
 and
 $$
\frac{\left[{\cal
N}_t(\Dl_{t-1})\right]^+ } {1+V(\Dl_{t-1})} = 
 \gm_t ~ \frac{\left[ 2\Dl_{t-1}^T  R_t(z^o+\Dl_{t-1})+ \gm_t E\left\{ \| \Psi_t(z^o+\Dl_{t-1}) \|^2 \mid {\cf}_{t-1} \right\} \right]^+ } 
 {1+\|\Dl_{t-1}\|^2} 
$$ 
  Since  $z^o+\Dl_{t-1}=Z_{t-1} \in U_{t-1}$,  \eqref{N+conv} 
follows from (C1). 
 $\diamondsuit $

\begin{corollary}\label{SqC} 
 Suppose that the conditions of Corollary \ref{SqS} hold and
 \begin{description}
 \item[(C2)]  for each $\ve\in (0, 1),$
$$
\sum_{t=1}^\infty \inf_u
\left[{\cal N}_t(u)\right]^-=\infty, \qquad P\mbox{-a.s.}$$
where
$$
{\cal N}_t(u)= 2u^T \gm_t R_t(z^o+u)+ \gm_t^2E\left\{ \| \Psi_t(z^o+u) \|^2 \mid {\cf}_{t-1} \right\} 
$$
and the   infimum is taken over the set 
$\{u: {\ve \le \|u\| \le {1/\ve}}; ~ {z^0+u \in U_{t-1}}\}.$
\end{description}
Then   $Z_t \to z^o \;\;
(P$-a.s.),  for any  initial value $Z_0$.
 \end{corollary}

 \noindent
 {\bf Proof.}     Let us show that the conditions of Lemma \ref{CL} are satisfied  with  $V(u)=u^Tu=\|u\|^2$ and  
 $\gamma_t(z)=\gamma_t {\bf  I}$.
  It follows from the proof of Corollary  \ref{SqS} that all the conditions  of  Lemma \ref{SL}  hold with    
  $V(u)=u^Tu$. Hence,  $\|Z_t-z^0\|$ converges and \eqref{N-conv} holds.  Since
 $$
  \inf_{\stackrel{ \|u\| \geq \ve}{z^0+u \in U_t}}\|u\|^2  \ge \ve^2, $$ 
  condition \eqref{VV}  
  also  trivially holds.  Finally,  (C) is a consequence of (C2).    
    $\diamondsuit $ 
    
\begin{corollary}\label{SC}   Suppose that  $Z_t$  is a  process  defined by 
\eqref{TSA}, \eqref{GTSA1}  and \eqref{GTSA2},   with an admissible for $z^0 \in  \mathbb{R}^m$  
 truncation sequence  $U_t$ and
\begin{description} 
 \item[(1)]
 $$
 (z-z^0)^T R_t(z) \le 0 ~~~~~\mbox{for any} ~~~~~~  z \in U_t,
 $$
  eventually;
    \item[(2)]
     $$
       \sup_{z \in U_{t-1}} ~ \frac {\|R_t(z)  \|^2}{1+\| z-z^o\|^2}\le r_t 
       $$    
 eventually,  where
 $$
  \sum_{t=1}^{\infty}   {r_{t}}{\gamma_t^2} <\infty, \qquad P\mbox{-a.s.}, 
  $$
    \item[(3)]
  $$
   \sup_{z \in U_{t-1}} ~ \frac {E \left\{ \|\ve_t(z)\|^2 \mid{\cf}_{t-1}\right\}}
  {1+\| z-z^o\|^2}\le e_t $$
 eventually,  where
 $$
  \sum_{t=1}^{\infty}   {e_{t}}{\gamma_t^2} <\infty, \qquad P\mbox{-a.s.}.
  $$
Then $\|Z_t-z^0\|$ converges  ($P$-a.s.) to a finite limit.
 \end{description}
   \end{corollary}
  {\bf Proof.}   Using condition (1),   
   $$
  \left[
2( z-z^o)^TR_t(z) +  \gm_t E \left\{ \|\Psi_t(z)\|^2 \mid{\cf}_{t-1}\right\} \right]^+ \le
\gm_t E \left\{ \|\Psi_t(z)\|^2 \mid{\cf}_{t-1}\right\} 
$$
eventually. 
Since   $E\left\{ \ve_t(z)\mid {\cf}_{t-1} \right\}=0$ and $R_t(z)$  is ${\cf}_{t-1}$-measurable, we have 
\begin{equation}\label{axali}
E\left\{ \| \Psi_t(z) \|^2 \mid {\cf}_{t-1} \right\} =
\|R_t(z) \|^2+E\left\{ \| \ve_t(z) \|^2 \mid {\cf}_{t-1} \right\}. 
\end{equation}
So, by conditions (2) and (3), the left hand side of \eqref{N+}   does not exceed  
$(r_t + e_t) \gamma_t$.  Hence   conditions   of  Corollary \ref{SqS}  hold 
with $q_t=(r_t + e_t) \gamma_t$ and the result follows.
 $\diamondsuit $
 

\begin{corollary}\label{CC}  
Suppose that the conditions of Corollary \ref{SC} are satisfied and  
\begin{description}
 \item[(CC)]   for each $\ve\in (0, 1),$
 \begin{equation} \label{INF}
 \inf_{\stackrel{ \ve \le \|z-z^o\| \le 1/\ve}{z\in U_{t-1}}}  -(z-z^0)^T R_t(z)> \nu_t
\end{equation}
eventually, where
$$
  \sum_{t=1}^{\infty}  {\nu_{t}}{\gamma_t} =\infty, \qquad P\mbox{-a.s.}
$$
\end{description}
Then $Z_t$ converges  ($P$-a.s.)  to $z^0$.
\end{corollary}

  {\bf Proof.} It follows from the poof of 
  Corollary \ref{SC} that conditions of 
   Corollary \ref{SqS} hold.  Let us prove that 
   (C2) of  Corollary  \ref{SqC} holds. Using   the obvious inequality $[a]^- \ge -a$,  we have
$$
\left[{\cal N}_t(u)\right]^-  \ge   - 2u^T \gm_t R(z^o+u)- \gm_t^2E\left\{ \| \Psi_t(z^o+u) \|^2 \mid {\cf}_{t-1} \right\}.
 $$
 Using \eqref{axali} and conditions (2) and (3)   of  Corollary \ref{SC},    and taking  the  supremum  of  the conditional expectation above  over the set  
 $\{u: {\ve \le \|u\| \le {1/\ve}}; ~ {z^0+u \in U_{t-1}}\}$,  we  obtain
 $$
 \sup \frac{E\left\{ \| \Psi_t(z^o+u) \|^2 \mid {\cf}_{t-1} \right\}}{1+\|u\|^2} ~(1+\|u\|^2)
 \le (r_t + e_t) (1+\|1/\ve\|^2).
  $$
 Then, by   \eqref{INF},  taking the   infimum over  the same set, 
$$
\inf \left[{\cal N}_t(u)\right]^-  \ge    2\gm_t \nu_{t} - \gm_t^2(r_t + e_t)(1+\|1/\ve\|^2).
 $$     
Condition  (C2) is  now immediate   from (CC) and    conditions (2) and (3) of Corollary \ref{SC}. Hence, by  Corollary  \ref{SqC}, 
 $Z_t$ converges  ($P$-a.s.)  to $z^0$. $\diamondsuit $
 
 \medskip
\begin{remark} 
Suppose that   $\ve_t$ is an  error term which does not depend on $z$ and denote 
  $$
\sigma_t^2= {E \left\{ \|\ve_t\|^2 \mid{\cf}_{t-1}\right\}}
$$
Then condition (3) holds if 
\begin{equation}\label{statefree}
  \sum_{t=1}^{\infty} \sigma_t^2 {\gamma_t^2} <\infty, \qquad P\mbox{-a.s.}.
  \end{equation}
  This shows that the requirement on the error terms are quite weak. In particular, the conditional variances  do not have to be bounded 
  w.r.t. t. 
  \end{remark}
  \begin{remark}\label{Comp}
As it was mentioned in  the introduction, our procedure is similar to the one considered in \cite{Andr}.
Let us compare these two  in the cases when the comparisons are possible. Hence,  
consider  truncations on  increasing   non-random sets,   non-random and homogeneous $R_t(u)=R(u)$, and scalar and 
state-independent $\gamma_t$ in Corollaries \ref{SC} and \ref{CC}.  
Also,  in  Theorem 2 of \cite{Andr}  take $\beta_n=0$.  
Then the resulting two sets of conditions are in fact  equivalent.  In particular,  in terms of notation in  \cite{Andr}, 
$$
a_n=\gamma_n, ~~~ \frac 1{c_n^2}=e_n, ~~~ M_n^2=r_n.
$$
Now it is clear that  conditions 2. and 3. in Theorem 2 of  \cite{Andr} are  equivalent to  (3) and (2)  respectively in Corollary \ref{SC}.  
Note  that although condition (CC) in \ref{CC} is formally more general than  condition 2. in Theorem 2 of \cite{Andr},
in any meaningful applications they are equivalent. 
  \end{remark}

\subsection{Examples}                   
 
 {\bf Example 1}   
Let  $l$  be  an odd integer and 
 $$
 R(z)=-(z-z^0)^l,
 $$                
$z , z^0 \in \mathbb{R}$. Consider a truncation sequence  $[-\alpha_t, \alpha_t]$, where $\alpha_t \to \infty$ is a  sequence of
 positive numbers.   Suppose  that  
 $$
   \sum_{t=1}^{\infty}   \gamma_{t}=\infty    ~~~ \mbox{and} ~~~ \sum_{t=1}^{\infty}   {\alpha_{t-1}^{2l}}~ {\gamma_t^2} <\infty.
$$
 Then, provided that the measurement errors satisfy \eqref{statefree}
 (or condition (3)  of Corollary \ref{SC}  in the case 
 of state-dependent errors),  the truncated procedure  
$$
Z_t=
\big[~
Z_{t-1}+\gm_t\left(R(Z_{t-1})+\ve_t\right)\big]_{-\alpha_t}^{\alpha_t},   \quad t=1,2,\dots
$$
 converges a.s. to $z^0$.
 
 Indeed,  condition (1) of Corollary \ref{SC} trivially holds. For large $t$'s,
 $$
   \sup_{z \in [-\alpha_{t-1,} \alpha_{t-1}]} ~ \frac {\|R(z)  \|^2}{1+\| z-z^o\|^2}\le  
\sup_{z \in [-\alpha_{t-1,} \alpha_{t-1}]} (z-z^0)^{2l}   \le 4^l \alpha_{t-1}^{2l}
 $$
  which implies  condition (2)  of Corollary \ref{SC}.   Condition (CC) of Corollary \ref{CC}    also trivially holds with $\nu_t=\ve^{l+1}$.

  For example, if the degree of the polynomial   is known to be $l$ (or at most $l$), and $\gamma_t=1/t$,  then one can take $\alpha_t=Ct^{\frac1{2l}-\delta}$, where $C$ and $\delta$ are some positive constants and $\delta < \frac{1}{2l}$.  One can also take a
  truncation sequence which is independent of $l$, e.g.,  $\alpha_t=C \log t$, where $C$ is a positive constant.

  \medskip

\noindent
{\bf Example 2 }
Let  $X_1,X_2,\ldots ,$  ~~   be i.i.d.~~ Gamma$(\theta,1), \;\;\ \theta >0.$
Then the  the common probability density function  is
$$
f(x,\theta)=\frac1{ {\bf{\Gamma}(\theta)} } x^{\theta-1}e^{-x}, \;\;\;
\theta >0, \;\; x >0,
$$
where ${\bf {\Gamma}(\theta)}$ is the Gamma function.
Then 
$$
\frac {f'(x,\theta)}{f(x,\theta)}={\log} x-
\underbrace{\frac{d}{d\theta}{\log} {\bf{\Gamma}(\theta)}}_
{{\log}' {\bf{\Gamma}(\theta)}},  ~~~~~~~~~
  {i} (\theta)=\underbrace{\frac{d^2}{d\theta^2}{{\log}} {\bf{\Gamma}(\theta)}}_
{{\log}'' {\bf{\Gamma}(\theta)}},
$$
where $ {i} (\theta)$ is the one-step Fisher  information.
Then a likelihood type  recursive estimation procedure  (see also \eqref{reciid}) can be defined as
 \begin{equation} \label{EstG}                      
\hat \theta_t=\left[\hat \theta_{t-1}+\frac1{t ~ {\log}'' {\boldsymbol{\Gamma}(\hat\theta_{t-1})}}\left(
\log X_t-{{\log}' {\boldsymbol{\Gamma}(\hat\theta_{t-1})}}\right)
\right]_{\alpha_t}^{\beta_t}, 
\qquad t=1,2,\dots
\end{equation}
where
$\alpha_t\downarrow  0$ and $\beta_t\uparrow \infty $ are sequences of positive numbers.

Everywhere in this example,  ${\cf}_{t}$ is the sigma algebra generated by $X_1, \dots, X_t$,  $P^\theta$ is the family of corresponding measures, and $\theta > 0$ is an arbitrary but fixed value of the parameter. 

Let us rewrite  \eqref{EstG}   in the form of the stochastic approximation, i.e.,
 \begin{equation} \label{SapG}                      
\hat \theta_t=\left[\hat \theta_{t-1}+\frac1{t} \left( R(\hat\theta_{t-1}) +\ve_t(\hat\theta_{t-1}) \right)
\right]_{\alpha_t}^{\beta_t}, 
\qquad t=1,2,\dots
\end{equation}
where (see Section \ref{App} for details)
$$
R(u)=R^\theta(u)=\frac1{{\log}'' {\boldsymbol{\Gamma}(u)}}E^\theta\{{\mbox{ln}} X_t-\log' {\bf {\Gamma}}(u)\}=
\frac1{{\log}'' {\boldsymbol{\Gamma}(u)}}\left(\log' {\bf {\Gamma}}(\theta)-
\log' {\bf {\Gamma}}(u)\right)
$$
and 
$$
\ve_t(u)=\frac1{{\log}'' {\boldsymbol{\Gamma}(u)}}\left(
\log X_t-{{\log}' {\boldsymbol{\Gamma}(u)}}\right) - R(u).
$$
Since  
$
E^\theta\left\{ \log X_t\mid {\cf}_{t-1} \right\}=E^\theta\left\{ \log X_t  \right\}=\log' {\bf {\Gamma}}(\theta)
$
and
$\hat\theta_{t-1}$  is ${\cf}_{t-1}$ - measurable,   we have
$
E^\theta\left\{ \ve_t (\hat\theta_{t-1})\mid {\cf}_{t-1} \right\}=0 
$
and hence   \eqref{GTSA1}  holds.  
Since $E^\theta\left\{ \log^2 X_t  \right\} < \infty$,   condition \eqref{GTSA2} can be checked in the similar way.
Obviously,  $R(\theta)=0$, and since  $\log' {\bf {\Gamma}}$ is   increasing (see, e.g.,  \cite{Wit}, 12.16),
condition {\bf (1)} of Corollary \ref{SC} holds with $z^0=\theta$.   Based on the well known properties of the logarithmic derivatives 
of the gamma function, it is not difficult to show  (see Section \ref{App})  that  if 
 \begin{equation} \label{albt}         \sum_{t=1}^\infty
\frac {\alpha_{t-1}^2} {t}= \infty   ~~~~~ \mbox{and} ~~~~~ \sum_{t=1}^{\infty} \frac{\log^2\alpha_{t-1}
+\log^2\beta_{t-1}}{t^2}  < \infty,
\end{equation}
then all the conditions of  Corollary \ref{SC} and \ref{CC}  hold and therefore, 
$\hat\theta_t$ is consistent, i.e.,
$$
\hat\theta_t\to \theta  ~~~~~~~ \mbox{as}~~~~~~~~~ t\to
\infty  ~~~~  \mbox{($P^\theta$-a.s.)}.
$$
For instance,  the sequences 
$$
{\alpha_t=C_1 ({{\log}} \; (t+2))^{-\frac12}} \;\;\; \mbox{and} \;\;\;
{\beta_t=C_2(t+2)}
$$
with  some positive constants $C_1$ and $C_2$, obviously satisfy  \eqref{albt}.

Note also, that since $\theta \in (0, \infty)$,  it may seem unnecessary to use the upper truncations
$\beta _t < \infty$. However,  without upper truncations (i.e. if $\beta _t =\infty$), the standard 
 restriction on  the  growth  does not hold.
   Also, with $\beta _t =\infty$   the procedure fails condition (2) of Corollary \ref{SC} (see \eqref{uperT}).
 
 \medskip
\noindent
{\bf Example 3} 
Consider an  AR(1) process
\begin{equation}\label{AR}
X_t=\theta X_{t-1}+\xi_t,
\end{equation}
where
${\xi}_t$ is a sequence of random variables with mean zero. Taking  
$$\Psi_t(z)= X_{t-1}\left(X_t-zX_{t-1}\right) 
$$
 $\gamma_t(z)=\gamma_t=\hat I_t=\hat I_0+\sum_{s=1}^t X_{t-1}^2,$  and $U_t= \mathbb{R}$, procedure 
 \eqref{TSA} reduces to the  recursive  least squares (LS)  estimator of $\theta$, i.e., 
                          \begin{eqnarray}\label{ArLsq}
&&\hat\theta_t=\hat\theta_{t-1}+\hat I_t^{-1}
X_{t-1}\left(X_t-\hat \theta_{t-1}X_{t-1}\right) ,\\
&&\hat I_t=\hat I_{t-1}+X_{t-1}^2, \qquad t=1,2,\dots \nonumber
\end{eqnarray}
where $\hat\theta_0$  and  $\hat I_0 > 0$ are any starting points.

For simplicity let us assume that ${\xi}_t$ is a sequence of i.i.d. r.v.'s with mean zero and variance $1$. 
Consistency  of   \eqref{ArLsq}  can be derived   from  our results for any  $ \theta  \in\mathbb{R}$ and
 without any further moment assumptions  on the innovation process $\xi_t$.  Indeed, 
 assume that $\theta$ is an arbitrary but fixed value of the parameter. 
Then, using  \eqref{AR},  we obtain 
$$
X_t-\hat \theta_{t-1}X_{t-1}=\xi_t+X_{t-1}(\theta-\hat \theta_{t-1}).
$$
and   \eqref{ArLsq}  can be rewritten as
        \begin{equation}\label{ArAS}
        \hat\theta_t=\hat\theta_{t-1}+\hat I_t^{-1}
        \left(X_{t-1}^2(\theta-\hat \theta_{t-1}) + X_{t-1}\xi_t\right). 
\end{equation}
So, \eqref{ArAS}  is a SA procedure with 
        \begin{equation}\label{RR}
        R_t(z)=X_{t-1}^2(\theta -z),
\end{equation}
  $\ve_t(z)=\ve_t=X_{t-1}\xi_t$, ~ 
 $\gm_t=\hat I_t^{-1}$ and $U_t=\mathbb{R}$.
Let us check condition (C1) of  Corrolary  \ref{SqS} with $z^0=\theta$ and  $U_t=\mathbb{R}$.   
Since   $E\left\{ \ve_t\mid {\cf}_{t-1} \right\}=0$ and $R_t(z)$  is ${\cf}_{t-1}$ measurable, 
 \eqref{GTSA1} and  \eqref{GTSA2}   trivially hold.  Also, 
        \begin{equation}\label{NNN}
        E\left\{ \| \Psi_t(z) \|^2 \mid {\cf}_{t-1} \right\} =
\|R_t(z) \|^2+E\left\{ \| \ve_t \|^2 \mid {\cf}_{t-1} \right\}
= X_{t-1}^4(\theta -z)^2+X_{t-1}^2,
\end{equation}
denoting  the expression in the  square brackets in  \eqref{N+}  by $w_t(z)$  (with $z^o=\theta$),we obtain  
  \begin{equation}\label{W1}
    w_t(z)=-2X_{t-1}^2(z-\theta)^2 +\hat I_t^{-1}X_{t-1}^4(\theta -z)^2+\hat I_t^{-1}X_{t-1}^2
\end{equation}
 \begin{equation}\label{W2}
 =-\delta X_{t-1}^2(z-\theta)^2 -X_{t-1}^2(z-\theta)^2
\left((2-\delta)- \hat I_t^{-1}X_{t-1}^2 \right)
+\hat I_t^{-1}X_{t-1}^2
\end{equation}
  for some $0< \delta<1$. Since $\hat I_t^{-1}X_{t-1}^2 \le 1$, 
  the positive part of the above expression does not exceed $\hat I_t^{-1}X_{t-1}^2$.
  This implies that   \eqref{N+} holds with $q_t= \hat I_t^{-1}X_{t-1}^2$.  Now, note that 
  if  $d_n$ is a  nondecreasing sequence  of positive numbers such that $d_t\to +\infty$
and  $ \tr d_t=d_t-d_{t-1},$ then
$\sum_{t=1}^\infty \tr d_t/d_t=+\infty $ and 
 $ \sum_{t=1}^\infty \tr d_t/d_t^{2} <+\infty.
$
So, for  $X_{t-1}^2=\tr \hat I_t$, since $ \hat I_t \to \infty$ for any $\theta\in \mathbb{R}$
 (see, e.g, Shiryayev \cite{Shir}, Ch.VII, $\S$5) , we have 
  \begin{equation}\label{II}
  \sum_{t=1}^{\infty}  \hat I_t^{-2} X_{t-1}^2 <\infty ~~~ \mbox{and} ~~~ 
\sum_{t=1}^{\infty}  \hat I_t^{-1}X_{t-1}^2=\infty.
 \end{equation}
  Hence, taking  $ q_t\gamma_t=  \hat I_t^{-2}X_{t-1}^2, $
 (C1) follows.  
  Therefore, $(\hat\theta_t-\theta)^2$ converges to a finite limit.  To show convergence to $\theta$,
  let us check condition (C2) of  of  Corrolary  \ref{SqC} with $z^0=\theta$ and  $U_t=\mathbb{R}$. 
  Using \eqref{RR} and \eqref{NNN}, we have
  $$
{\cal N}_t(u)=-2\hat I_t^{-1}X_{t-1}^2u^2 +\hat I_t^{-2}X_{t-1}^4u^2+\hat I_t^{-2}X_{t-1}^2=
\hat I_t^{-1} w_t(\theta+u),
$$   
 where $w_t$ is defined in  \eqref{W1}. Since the middle term in   \eqref{W2} is non-positive, 
 using  the obvious inequality $[a]^- \ge -a$, we can write
 $$
\left[{\cal N}_t(u)\right]^-\ge   \delta \hat I_t^{-1}X_{t-1}^2 u^2 -\hat I_t^{-2}X_{t-1}^2,
$$
and
$$
 \sum_{t=1}^\infty
\inf_{\ve \le |u| \le {1/\ve}} \left[ {\cal N}_t(u) \right]^- =\infty
$$
now follows from \eqref{II}. So, by Corollary \ref{SqS},  $\hat\theta_t\to \theta$   ($P^\theta-$ a.s.).

 Note that the convergence of the LS estimator  is well known  under these assumptions.
 (see e.g., \cite{Shir}, Ch.VII, $\S$5).  This example is presented  to demonstrate 
 that the assumptions  made here are minimal.  That is, in well know model cases,  the results of the paper do not assume any additional restrictions.

\section{Appendix}\label{App}

We will need the following properties of  the Gamma function  (see, e.g.,  \cite{Wit}, 12.16).
${\log}' {\boldsymbol{\Gamma}}$ is  increasing, 
    ${\log}'' {\boldsymbol{\Gamma}}$  is decreasing and continuous, and 
  $$                      
{\log}'' {\boldsymbol{\Gamma}}(x)=
\frac 1{x^2}+\sum_{n=1}^\infty \frac1{(x+n)^2}.
$$
The latter implies that
\begin{equation}\label{Log''G1}
{\log}'' {\boldsymbol{\Gamma}}(x) \le 
\frac 1{x^2}+\sum_{n=1}^\infty \int_{n-1}^n \frac{dz}{(x+z)^2}
=\frac 1{x^2}+\frac 1x=\frac {1+x}{x^2}
\end{equation}
and
\begin{equation}\label{Log''G2}
{\log}'' {\boldsymbol{\Gamma}}(x) \ge
\sum_{n=0}^\infty \int_n^{n+1} \frac{dz}{(x+z)^2}=\frac 1x.
\end{equation}
Also (see \cite{Cram},  12.5.4), 
\begin{equation} \label{Log'G}
{\log}' {\boldsymbol{\Gamma}}(x) \le {\mbox{ln}} (x).
\end{equation}
Then,
\begin{equation}\label{expect}
E^\theta\left\{\log  X_1 \right\}= {\log}' {\boldsymbol{\Gamma}}(\theta)            ~~~~ \mbox{and} ~~~~
E^\theta\left\{ \left(\log X_1\right)^2 \right\}={\log}'' {\boldsymbol{\Gamma}}(\theta) +
 \left({\log}' {\boldsymbol{\Gamma}}(\theta)\right)^2
  \end{equation}
and
$$
E^\theta\left\{ \left(\log X_1 -{\log}' {\boldsymbol{\Gamma}}(\theta)\right)^2 \right\}={\log}'' {\boldsymbol{\Gamma}}(\theta). 
$$
Let us show that the conditions of Corollary \ref{SC} hold.  Since 
$$
\Psi_t(u)=\frac 1{ {\log}'' {\boldsymbol{\Gamma}(u)}}
\left(
\log X_t-{{\log}' {\boldsymbol{\Gamma}(u)}}\right),
$$
using  \eqref{expect} and \eqref{Log''G2}  we obtain
\begin{equation}\label{Sq}
\frac {E \left\{ \|\Psi_t(u)\|^2 \mid{\cf}_{t-1}\right\}}{1+\| u-\theta\|^2}=\frac{ {\log}'' {\boldsymbol{\Gamma}(\theta)}
+\left(\log' {\bf {\Gamma}}(\theta)-
\log' {\bf {\Gamma}}(u)\right)^2}   {({\log}'' {\boldsymbol{\Gamma}(u))^2}  ({1+\| u-\theta\|^2})}
\end{equation}
$$
\le 
\frac{u^2} {1+ (u-\theta)^2} \left( {\log}'' {\boldsymbol{\Gamma}(\theta)}
+\left(\log' {\bf {\Gamma}}(\theta)-
\log' {\bf {\Gamma}}(u)\right)^2\right).
$$
Now,  ${u^2}/({1+ (u-\theta)^2})  \le C$. Here and further on  in this subsection, $C$  denotes various constants which may depend on $\theta$. So, using \eqref{Log'G}    we obtain
$$
\frac {E \left\{ \|\Psi_t(u)\|^2 \mid{\cf}_{t-1}\right\}}{1+\| u-\theta\|^2}
\le C\left(  {\log}'' {\boldsymbol{\Gamma}(\theta)}
+  \log' {\bf {\Gamma}}(\theta)^2+
\log' {\bf {\Gamma}}(u)^2\right)
\le C(1+\log^2(u)).
$$
For large $t$'s, since $ \alpha_t < 1 <  \beta_t $, we have
$$ 
 \sup_{u \in [\alpha_t, \beta_t]} ~\log^2(u) 
 \le  \left\{ \sup_{\alpha_t\le u < 1}  {\log}^2(u)+\right.
\left.  \sup_{1 <u \le \beta_t}  {\log}^2(u)\right\} \le
 {\log}^2 \alpha_t+ {\log}^2 \beta_t.
$$
  Condition (2) of Corollary \ref{SC} is now immediate  from the second part of \eqref{albt}.
 It remains to check that (CC) of Corollary \ref{CC} holds. Indeed,
 $$
 -(u-\theta) R(u)= 
 \frac{(u-\theta)\left(\log' {\bf {\Gamma}}(u)-
\log' {\bf {\Gamma}}(\theta)\right)}
 {{\log}'' {\boldsymbol{\Gamma}(u)}}.
$$ 
Since ${\log}' {\boldsymbol{\Gamma}}$ is  increasing and   
    ${\log}'' {\boldsymbol{\Gamma}}$  is decreasing and continuous, we have that for each $\ve\in (0, 1),$
 \begin{equation} 
 \inf_{\stackrel{ \ve \le \|u-\theta\| \le 1/\ve}{u\in U_{t-1}}} 
  -(u-\theta) R(u)\ge 
 \frac{ \inf_{ \ve \le \|u-\theta\| \le 1/\ve}  \left(\log' {\bf {\Gamma}}(u)-\log' {\bf {\Gamma}}(\theta)\right)
 (u-\theta)}
 { \sup_{u\in U_{t-1}} {\log}'' {\boldsymbol{\Gamma}(u)}}
\ge \frac C{ {\log}'' {\boldsymbol{\Gamma}(\alpha_{t-1})}}
\end{equation}
where $C$ is a constant that my depend on $\ve$ and $\theta$. Since  $\alpha_{t-1}<1$ for large $t$'s, 
it follows \eqref{Log''G1}  that  $1/{ {\log}'' {\boldsymbol{\Gamma}(\alpha_{t-1})}}\ge \alpha_{t-1}^2/2$.
  Condition (CC) of Corollary \ref{CC} is now immediate  from the first part of \eqref{albt}.
  
  Note that  with  $\beta _t =\infty$   the procedure fails
 condition (2) of Corollary \ref{SC}. Indeed, \eqref{Sq} and \eqref{Log''G1} 
implies that
\begin{equation}\label{uperT}
\sup_{\alpha_t\le u} \frac {E \left\{ \Psi_t^2(u) \mid{\cf}_{t-1}\right\}}{1+( u-\theta)^2}\ge
\sup_{\alpha_t\le u}\frac{ \left\{{\log}'' {\boldsymbol{\Gamma}(\theta)}+\left(\log' {\bf {\Gamma}}(\theta)-
\log' {\bf {\Gamma}}(u)\right)^2 \right \} {u^4}} { {(1+u)^2}{(1+( u-\theta)^2})}=\infty
\end{equation}


\begin{thebibliography}{9}
%
 \bibitem{Andr} \textsc{Andrad\'ottir,} S.  (1995).       
A stochastic approximation algorithm with varying bounds. {\em Operations Research}
  {\bf 43}, 6, 1037--1048.
   %
%
 \bibitem{AMP} \textsc{Andrieu,} C., \textsc{Moulines,} E. and \textsc{Priouret,} P.  (2005).       
 Stability of stochastic approximation under verifiable conditions. {\em SIAM J. Control Optim.}
  {\bf 44},  283--312.
   %
 \bibitem{Ben} 
 \textsc{Benveniste,} A, \textsc{Metivier,} M. and   \textsc{Priouret ,} P. (1990).
 {\em Adaptive Algorithms and Stochastic Approximation}. Berlin and New
York: Springer-Verlag.
 %
 \bibitem{Bor} \textsc{Borkar,} V. S. (2008).
{\em Stochastic approximation: A Dynamical  Systems Viewpoint.}  Cambridge University Press.
%
\bibitem{Chen1}  \textsc{Chen,} H.,  \textsc{Guo,} L. and  \textsc{Gao,} A. (1987). Convergence and robustness of the Robbins-Monro algorithm truncated at randomly varying bounds. {\em Stochastic Processes 
Appl.}  {\bf 27},  217Ð231.
%
\bibitem{Chen2}  \textsc{Chen,} H. and  \textsc{Zhu,} Y.-M. (1986). Stochastic approximation procedures with 
randomly
varying truncations. {\em Scientia Sinica 1} {\bf  29},  914Ð926.
%
 \bibitem{Cam}
     \textsc{Campbell, K.} (1982). Recursive computation of
     M-estimates for the parameters of a finite autoregressive
     process. {\it Ann. Statist.} {\bf 10}, 442-453.
%
 \bibitem{Cram}  \textsc{Cramer,} H. (1946).
    {\em Mathematical Methods of Statistics}. Princeton University Press,
          Princeton.
 \bibitem{Eng}
 \textsc{{ Englund,} J.-E.,   {Holst,} U., and {Ruppert,} D.} (1989).
  Recursive estimators for
         stationary, strong mixing processes -- a representation
         theorem and asymptotic distributions.
         {\it Stochastic Processes Appl.}  {\bf 31}, 203--222.
          %
\bibitem{Fab}
\textsc{Fabian, V.} (1978).  On asymptotically  efficient
         re\-cur\-sive es\-ti\-ma\-tion.
         {\it Ann. Statist.} \textbf {6}, 854-867.
%
\bibitem{Gu}
\textsc{Gu,} M.G. and   \textsc{Li,} S. (1998).  A stochastic approximation algorithm for
maximum-likelihoood estimation with incomplete data.
         {\it The Canadian Journal of Statistics} \textbf {26}, 567-582.
\bibitem{Kall}
 \textsc{ Kallenberg, O.}    (1997).
  {\it Foundations of Modern Probability}. Nauka, Moscow.
\bibitem{Khas}
 \textsc{ Khas'minskii, R.Z., Nevelson, M.B.}    (1972).
  {\it Stochastic Approximation and
          Recursive Estimation}. Springer-Verlag, New-York.
 \bibitem{Kush1}
 \textsc{Kushner,}  H. (2010).  Stochastic approximation: a survey.
 {\em Wiley Interdisciplinary Reviews: Computational Statistics}
{\bf 2}, 6,   87--96.   
\bibitem{Kush2}
 \textsc{Kushner,}  H. and \textsc{Yin,} G. (1997). {\em Stochastic Approximation Algorithms and Applications.
Applications of Mathematics}.  Springer-Verlag, New-York.         
          %
\bibitem{Lai}
\textsc{Lai,} T.L.  (2003).  Stochastic approximation.
         {\it Ann. Statist.} \textbf {31}, 391-406.
         
             \bibitem{Lel}
\textsc{Lelong,} J.  (2008).  Almost sure convergence  of randomly truncated stochastic algorithms under verifiable
conditions.
         {\it Statistics $\&$ probability Letters.}  \textbf {28}, 2632Ð-2636.       
%
\bibitem{Ljun2}
  \textsc{{Ljung,}} L. and \textsc{{Soderstrom,}} T. (1987). {\it Theory and
Practice of Recursive Identification,} MIT Press.
%
\bibitem{Pol}
  \textsc{Polyak,} B. T. and  \textsc{Tsypkin,} Ya. Z.   (1980). 
  {Robust identification.}  {\em Automatica} 16, 53--69 
  %
\bibitem{RM}
  \textsc{{Robbins,}} H. and \textsc{{Monro,}}  S.  (1951) A stochastic approximation method,
         {\it Ann. Statist.}  {\bf 22},  400--407.
\bibitem{Rob2}
    \textsc{{Robbins,} H.  and {Siegmund,}  D.} (1971). A convergence theorem for
          nonnegative almost supermartingales and some applications.
         {\it Optimizing  Methods in Statistics}. ed. J.S. Rustagi
          Academic Press, New York,  233--257.
\bibitem{Sakr} \textsc{Sakrison,} D.J.  (1965). Efficient recursive estimation; application to
        estimating the parameters of a covariance function.
        {\em Internat. J. Engrg. Sci.}  {\bf 3}, 461--483.
  %
        \bibitem{Shar0}
     \textsc{{Sharia,} T.} (1997).
Truncated recursive estimation procedures,
          {\it Proc. A. Razmadze Math. Inst.} {\bf  115}, 149--159.
\bibitem{Shar}
     \textsc{{Sharia,} T.} (1998). On the recursive parameter estimation for the
          general discrete time statistical model. {\it
          Stochastic Processes Appl.}  {\bf 73}, {\bf 2}, 151--172.
\bibitem{Shar1}
 \textsc{Sharia, T.} (2008). Recursive parameter estimation: Convergence.
{\it Statistical Inference for Stochastic Processes}.  {\bf 11},  2,    pp. 157 -- 175.
%
\bibitem{Shar2}
  \textsc{Sharia, T.}   (2007). Rate of  convergence in recursive parameter
estimation procedures. {\it Georgian Mathematical Journal}.  {\bf 14}, {4}, pp. 721--736.
%
\bibitem{Shar3}
 \textsc{Sharia, T.}   (2010). Recursive  parameter estimation: Asymptotic expansion. {\it The Annals of The Institute
 of Statistical Mathematics} {\bf 62}  2, 343-362.
  %
\bibitem{Shar4}  \textsc{Sharia, T.}   (2010). Efficient On-Line Estimation  of Autoregressive Parameters.
 {\it  Mathematical Methods of Statistics}. {\bf 19}, 2, 163-186.
%
 \bibitem{Shir}
  \textsc{ Shiryayev, A.N.} (1984). {\it Probability,}
           Springer-Verlag, New York.
           %
  \bibitem{Tadic1}
  \textsc{Tadic,}                                                                                                                                                                                                                                           V.   (1997) Stochastic gradient with random truncations, {\em European J. of Operational Research}, {\bf 101}, pp. 261--284.
  %
   \bibitem{Tadic2}
  \textsc{Tadic,} V.    (1998) Stochastic approximations with random truncations, state dependent noise and discontinuous dynamics, {\em Stochastics and Stochastics reports}. {\bf 64},  pp. 283--326.        
 \bibitem{Wit}    \textsc{ Whittaker}, E. \textsc{ Watson,} G.  (1927). {\em A Course of Modern Analysis}.
         Cambridge University Press, Cambridge.       
\end{thebibliography}
                 \end{document}